\documentclass[twocolumn]{aastex631}

\usepackage{amsmath} 
\usepackage{booktabs}
\usepackage{multirow}
\usepackage{graphicx}

\shorttitle{HAT-P-26b}
\shortauthors{Ramos Rosado et al.}

\begin{document}

\title{HST SHEL: Revealing Haze and Confirming Elevated Metallicity in the Warm Neptune HAT-P-26b}

\author[0000-0002-1056-3144]{Lakeisha M. Ramos Rosado}
\affiliation{William H. Miller III Department of Physics and Astronomy, Johns Hopkins University, 3400 N. Charles Street, Baltimore, MD 21218, USA}
\correspondingauthor{Lakeisha M. Ramos Rosado}
\email{lramosr1@jhu.edu}

\author[0000-0001-6050-7645]{David K. Sing}
\affiliation{William H. Miller III Department of Physics and Astronomy, Johns Hopkins University, 3400 N. Charles Street, Baltimore, MD 21218, USA}
\affiliation{Department of Earth and Planetary Sciences, Johns Hopkins University, 3400 N. Charles Street, Baltimore, MD 21218, USA}

\author[0000-0002-0832-710X]{Natalie H. Allen}
\affiliation{William H. Miller III Department of Physics and Astronomy, Johns Hopkins University, 3400 N. Charles Street, Baltimore, MD 21218, USA}

\author[0000-0003-4328-3867]{Hannah R. Wakeford}
\affiliation{School of Physics, University of Bristol, HH Wills Physics Laboratory, Tyndall Avenue, Bristol BS8 1TL, UK}

\author[0000-0003-3204-8183]{Mercedes L\'opez-Morales}
\affiliation{Space Telescope Science Institute, 3700 San Martin Drive, Baltimore, MD 21218, USA}

\author[0000-0002-6500-3574]{Nikolay K. Nikolov}
\affiliation{Space Telescope Science Institute, 3700 San Martin Drive, Baltimore, MD 21218, USA}

\author[0000-0002-7352-7941]{Kevin B. Stevenson}
\affiliation{Johns Hopkins APL, 11100 Johns Hopkins Rd, Laurel, MD 20723, USA}

\author[0000-0003-4157-832X]{Munazza K. Alam}
\affiliation{Space Telescope Science Institute, 3700 San Martin Drive, Baltimore, MD 21218, USA}

\author[0000-0001-5442-1300]{Thomas M. Evans-Soma}
\affiliation{School of Information and Physical Sciences, University of Newcastle, Callaghan, NSW, Australia}
\affiliation{Max Planck Institute for Astronomy, K\"{o}nigstuhl 17, D-69117 Heidelberg, Germany}
 
\begin{abstract}

We present a new and extended transmission spectrum of the warm Neptune HAT-P-26b spanning wavelengths between 0.29 $-$ 5.0 $\mu$m. This spectrum is derived from new HST STIS G430L observations from the PanCET program, a reanalysis of the previously published HST STIS G750L data, along with the previously published HST WFC3 IR G102 and G141 data, and the two Spitzer IRAC  photometric points at 3.6 and 4.5 $\mu$m. We present this analysis as part of the Sculpting Hubble’s Exoplanet Legacy (SHEL) program, where the goals are to analyze all HST archival observations of transiting exoplanets using a uniform and homogeneous reduction technique. With the new wavelength coverage, we identify a scattering slope that is weaker than Rayleigh scattering and is best-matched by models incorporating a haze-only scenario. Our retrieval analysis reveals an atmospheric metallicity of 15$_{-8}^{+22}$$\times$ solar which suggests that HAT-P-26b may have formed further out in the protoplanetary disk, in a region rich in hydrogen and helium but with fewer heavy elements, and later migrated inward. This super-solar metallicity places HAT-P-26b below the mass-metallicity trend of the solar system. Looking ahead, recent observations from JWST NIRISS/SOSS and NIRSpec/G495H will provide critical, high-precision data that extend the spectral coverage into the infrared to further constrain the atmospheric composition and structure of HAT-P-26b. These observations have the potential to confirm or refine the metallicity and haze scenario presented here, offering unprecedented insights into the atmospheric properties of warm Neptunes and the processes governing their formation and migration histories.

\end{abstract}


\section{Introduction} \label{sec:intro} 

Over the past two decades, transmission spectroscopy observations with the \textit{Hubble Space Telescope} (HST) have proven highly successful in exploring the atmospheres of giant planets, resulting in the detection of multiple species. Some notable highlights from HST include detection of atomic species \citep[e.g.,][]{Charbonneau_2002,Vidal-Madjar_2003,Sing_2011,Huitson_2013,Nikolov_2014,Ehrenreich_2015,Sing_2016,Alam_2018,Sing_2019,Carter_2020,Alam_2021,Nikolov_2022} as well as clouds and hazes \citep{Lecavelier_2008,Bean_2010,Sing_2013,Kreidberg_2014,Sing_2016,Kirk_2017,Wakeford_2017_1,Alam_2018,Benneke_2019,Gao_2020,Rathcke_2023,Changeat_2024} and molecular detections \citep[e.g.,][]{Deming_2013, Wakeford_2013,Kreidberg_2015,Wakeford_2017_2, Kilpatrick_2018, Wakeford_2018}.

In addition, observational campaigns using ground-based telescopes at low resolution are expanding the number of giant planets with atmospheric characterization for example surveys using the \textit{Very Large Telescope} (VLT) FOcal Reducer and Spectrograph (FORS2) \citep[e.g.,][]{Bean_2010, Nikolov_2016, Gibson_2017,Nikolov_2018, Carter_2020}, the Arizona-CfA-Católica Exoplanet Spectroscopy Survey (ACCESS) \citep[e.g.,][]{Rackham_2017,Bixel_2019,Espinoza_2019, McGruder_2020,Kirk_2021,Weaver_2020,Weaver_2021, Allen_2022,McGruder_2022, McGruder_2023} and The Low Resolution Ground-Based Exoplanet Atmosphere Survey using Transmission Spectroscopy (LRG-BEASTS) \citep[e.g.,][]{Kirk_2018,Kirk_2019, Alderson_2020, Kirk_2021,Ahrer_2022, Ahrer_2023}.

Most of the results mentioned above correspond to atmospheric observations of Jupiter to Saturn-sized planets; however, some involve smaller, Neptune-sized planets that have yielded interesting findings (e.g., HAT-P-11b: \citealt{Deming_2011,Fraine_2014,Huber_2017}; GJ-436b: \citealt{Ehrenreich_2015,Bourrier_2016};
HAT-P-26b: \citealt{Stevenson_2016,Wakeford_2017_2,MacDonald_2019}).

Transmission spectra in optical and near-ultraviolet (NUV) wavelengths are particularly important because they enable studies of condensation clouds and photochemical hazes in the atmospheres of exoplanets. Physical processes like Rayleigh or Mie scattering produced by aerosols result in a continuum slope at these short wavelengths which can be used to understand the cloud composition and constrain haze properties \citep{Wakeford_2017_3,Alam_2020,Wong_2020}. Clouds and hazes are critical elements of planetary atmospheres, playing a pivotal role in controlling radiative processes such as transmission, reflection, and emission  \citep{Gao_2021}. In the context of transmission, clouds can absorb light which increases the opacity of the atmosphere at high altitudes preventing measurements deeper down in the atmosphere, significantly limiting our ability to accurately interpret exoplanet spectra. This effect is particularly notable in the water vapor (H$_2$O) feature at $\sim$ 1.4 $\micron$ \citep[e.g.,][]{Sing_2016}. The combination of observations in the optical and near-infrared can provide insights into a planet's metallicity through H$_2$O abundances and offer constraints on the opacity of clouds if they are present \citep{Gao_2020}. Constraining H$_2$O provides essential clues for identifying planet formation scenarios \citep[e.g.,][]{Oberg_2011,Mordasini_2016}. 

In the Solar System, the metallicity of giant planets is measured using CH$_4$ abundances and is expressed as the ratio C/H. For the gas giants, Jupiter and Saturn, C/H is $\sim$ 4$\times$ solar and $\sim$ 10$\times$ solar, respectively \citep{Atreya_2016}. In the case of the ice giants, Neptune and Uranus, they are $\sim$ 80$\times$ solar \citep{Karkoschka_2011,Sromovsky_2011}. This trend is consistent with the core accretion model of planet formation \citep{Pollack_1996}. For exoplanets, water absorption features have been widely used as a proxy for metallicity measurements \citep[e.g.,][]{Kreidberg_2014_2,Fraine_2014,Wakeford_2018}, expressed as the ratio O/H. However, studies of exoplanets with lower masses, including Neptune-like planets and super-Earths have shown to be challenging to constrain. \cite{Fraine_2014} reported the first detection of H$_2$O in the exo-Neptune HAT-P-11b with a derived metallicity ranging from 1 to 700$\times$ solar; thereafter, \cite{Wakeford_2017_2} published a well-constrained metallicity for HAT-P-26b with an uncertainty of slightly more than one order of magnitude. HST's coverage in the blue-optical wavelength range can help to disentangle the degeneracy between clouds and the water feature, providing constraints for clouds and metallicity measurements \citep{Fairman_2024}. 

The focus of this study is the warm Neptune-sized exoplanet HAT-P-26b discovered in 2011 \citep{Hartman_2011}. This planet lies in a 4.23 day orbit with R$_p$ = 6.33 R$_\oplus$ and M$_p$ = 18.6 M$_\oplus$ and a low gravity log g$_p$ = 2.65 cm s$^{-2}$. The low gravity in combination with the planet's temperature (T$_{eq}$ = 1001 K) contributes to a large atmospheric scale height ($\sim$804 km), making it favorable for characterization studies using transmission spectroscopy. HAT-P-26b has been widely studied through different facilities, first using the Spitzer Space Telescope \citep{Fazio_2004}, from ground-based telescopes \citep{Stevenson_2016,Vissapragada_2022, A-thano_2023}, with HST \citep{Wakeford_2017_2} and analytically \citep{MacDonald_2019}. This target is the first exo-Neptune with a well-constrained metallicity, O/H = 4.8$_{-4.0}^{+21.5}$$\times$ solar. \cite{Wakeford_2017_2} found that the metallicity of this planet falls below the trend in mass-metallicity observed in giant planets from the Solar System, and smaller than what is expected for a planet with its mass from core-accretion scenarios \citep{Fortney_2013}. Furthermore, \cite{MacDonald_2019} did a comprehensive atmospheric retrieval analysis, making use of all available observations, to understand the atmospheric properties of HAT-P-26b. They report an updated value of metallicity O/H = 18.1$_{-11.3}^{+25.9}$$\times$ solar, suggesting a formation scenario with planetesimal accretion, which is still below that of Neptune. These results raise questions about HAT-P-26b's formation and evolution process. Motivated by these studies, we add a new optical transmission spectrum to all the previous observations of this planet and provide an extended transmission spectrum from the optical to infrared with a retrieval analysis of its atmospheric properties. 

In this paper, we present a new optical transmission spectrum of the warm Neptune HAT-P-26b, observed as part of the HST Panchromatic Comparative Exoplanet Treasury (PanCET) program (GO-14767, PIs: D. Sing and M. López-Morales). The scientific objectives of PanCET are to deliver a consistent and statistically significant comparative study of 20 exoplanet atmospheres, focusing on clouds, hazes, and chemical compositions across the ultraviolet (UV), optical and infrared (IR) wavelengths. This program provides unique observations in the UV and blue-optical wavelengths which are not accessible with the \textit{James Webb Space Telescope} (JWST). The analysis of this project was carried out in conjunction with the Sculpting Hubble’s Exoplanet Legacy (SHEL) program (HST GO-16634, PI: D. Sing). Where the goals of the program are to analyze all HST archival observations of transiting exoplanets using a uniform and homogeneous reduction technique. The structure of this paper is as follows: \autoref{sec:Observations} describes the observations and data reduction process, \autoref{sec:Analysis} details the systematics detrending and analysis of the transit light curves, in \autoref{sec:Results} we present the results along with a discussion of them in \autoref{sec:Discussion} and present the conclusions in \autoref{sec:Conclusion}. 
\vspace{0.43cm}

\begin{table}[h]
\caption{Stellar and Planetary Parameters of the HAT-P-26 system.}
\hspace{-1.6cm}
\centering
\begin{tabular}{c c c c}
\hline\hline
Stellar Parameters & Value & Ref.\\
\hline
Radius & 0.788$_{-0.043}^{+0.098}$ $R_\sun$ & [1]\\
Mass & 0.816 $\pm$ 0.033 $M_\sun$ & [1]\\ 
Density ($\rho_*$) & 1774 $\pm$ 108 kg/m$^3$ & [3]\\
Eff. temperature ($T_{eff}$) & 5011 $\pm$ 55 K & [2]\\
\hline
Planetary Parameters & Value &
Ref.\\
\hline
Radius & 0.565$_{-0.032}^{+0.072}$ $R_J$ & [1]\\
Mass & 0.059 $\pm$ 0.007 $M_J$ & [1]\\
Orbital period ($P$) & 4.234516 $\pm$ 0.000015 days & [1]\\
Scaled semi-major axis ($a/R_*$) & 11.89 $\pm$ 0.417 & [3]\\
Transit center time ($T_0$) & 2456901.059458 BJD & [4]\\
Impact parameter ($b$) & 0.395$\pm$ 0.116 & [3]\\
Eccentricity ($e$) & 0.124 $\pm$ 0.060 & [1]\\ 
Inclination ($i$) & 88.09  $\pm$ 0.553 deg & [3]\\
Surface gravity (log $g_{p}$) & 2.65$_{-0.10}^{+0.08}$ cm/s$^2$ & [1]\\
Argument of periastron ($\omega$) & 54 $\pm$ 165 deg & [1]\\
Eq. Temperature ($T_{eq}$) & 1001$_{-37}^{+66}$ K & [1]\\
Distance ($a$) & 0.0479 $\pm$ 0.0006 AU & [1]\\
\hline
\end{tabular}
\vspace{0.2cm}
{\small [1] \cite{Hartman_2011}\\
\small [2] \cite{Mortier_2013}\\
\small [3] \cite{Wakeford_2017_2}}\\
\small [4] this work
\label{table 1}
\end{table}
\vspace{-0.5cm}

\section{Observations and Data Reduction} \label{sec:Observations}

\subsection{Observations}

We observed two transits of HAT-P-26b as part of the HST PanCET program, both taken with the Space Telescope Imaging Spectrograph (STIS) G430L grating. Each transit was observed in a single visit during 2018 March 27 (visit 14) and 2019 June 14 (visit 13). The G430L grating covers the wavelength region from 2892  \text{\AA} to 5700 \text{\AA}, providing a low-resolution of R = $\lambda/\Delta\lambda$ = 500. The visits were planned to include the transit event in the third orbit, ensuring sufficient out-of-transit baseline flux and good coverage between the second and third contact. HST orbits the Earth every $\sim$ 96 minutes, with data collection halted during each orbit when HST is occulted by the Earth. This approach yielded five consecutive orbits, resulting in a total of 48 spectra per visit per instrument. Of these, we used 43, discarding the first exposure of each HST orbit, as it has historically shown lower fluxes than the subsequent exposures \citep{Brown_2001,Sing_2011,Sing_2019}. Each spectrum had an exposure time of 253 seconds, they were taken with the 52 $\times$ 2 arcsec$^2$ slit to minimize slit losses, and using a reduced sub-array of 1024$\times$128\, pixels to reduce readout times. 

\subsection{Data Reduction}

We reduced the STIS G430L spectra following the methodology described in \cite{Allen_2024}, which utilizes the pipeline from the SHEL program. This approach is briefly summarized here. We start with the flat-fielded products from MAST ({\tt *\_flt.fits}) as well as the engineering jitter files ({\tt *\_jit.fits}) and extract the respective time-series exposures. At this stage, we start the data cleaning processes. Initially, we use the cleaning function to identify pixels with a poor data quality flag in STIS. Subsequently, we use the technique called difference images to remove cosmic rays as described in \cite{Nikolov_2014}, which appear as an excess in flux. The next step is to find and correct for hot and cold pixels, that will not be detected using in the previous steps, using a spline-fitting method. Lastly, implementing the same spline-fitting technique, we further investigate any potentially faulty pixels that may have been overlooked. Once we have our data cleaned, we proceed with the spectral extraction method using two different techniques. We use optimal extraction \citep{Marsh_1989} as implemented in \citet{Brahm_2017}. This approach uses a series of polynomials to model the spectral profile. The method assigns a weight to each pixel in the science frame, facilitating a weighted extraction of the spectrum. We apply an aperture size of 15 pixels and we compare it to the ``classical" method of a spectral extraction which uses a box with a 13-pixel aperture width around the central trace \citep{Huitson_2013,Sing_2013,Wakeford_2017_2}. We chose to proceed with the optimal extraction method as no significant differences were observed between the resulting spectra and the optimal extraction was strongly justified in \cite{Allen_2024}. For more details of each step in the data reduction process, see \cite{Allen_2024}. To ensure consistent analysis across the spectrum, a goal of the SHEL program, we reanalyze the STIS G750L observations presented in \cite{Wakeford_2017_2} implementing the same analysis approach detailed above.

\section{Analysis} \label{sec:Analysis}

\subsection{Systematics Detrending} \label{sec:systematics}

Time-series observations taken with HST STIS are highly affected by instrument-related systematics. The main effect is due to the well-known thermal breathing of HST, a process in which the telescope warms up and cools down during its orbital cycle. This in turn causes the point-spread function (PSF) to change periodically and produces photometric changes in the light curves \citep{Brown_2001} which are correlated to HST's orbital phase. These systematic errors in the spectrophotometry have limited the precision achieved. To model the systematics, we follow a procedure similar to \cite{Sing_2011,Sing_2019} and as implemented in the SHEL pipeline as described in \cite{Allen_2024}.

We utilize six jitter vectors that demonstrated the strongest correlation with the data: {\it V2\_roll, V3\_roll, RA, DEC, Latitude} and {\it Longitude}. We also include three detrending parameters related to the spectral trace movement, one being the start of the trace in the y-value and two coefficients coming from the trace fitting. Lastly, we include a fourth order polynomial of HST's orbital phase. To decorrelate these detrending vectors, we first normalize each measured quantity by subtracting the mean value and then dividing by the standard deviation.

\begin{figure*} [ht]
    \centering
    \includegraphics[width=1\textwidth]{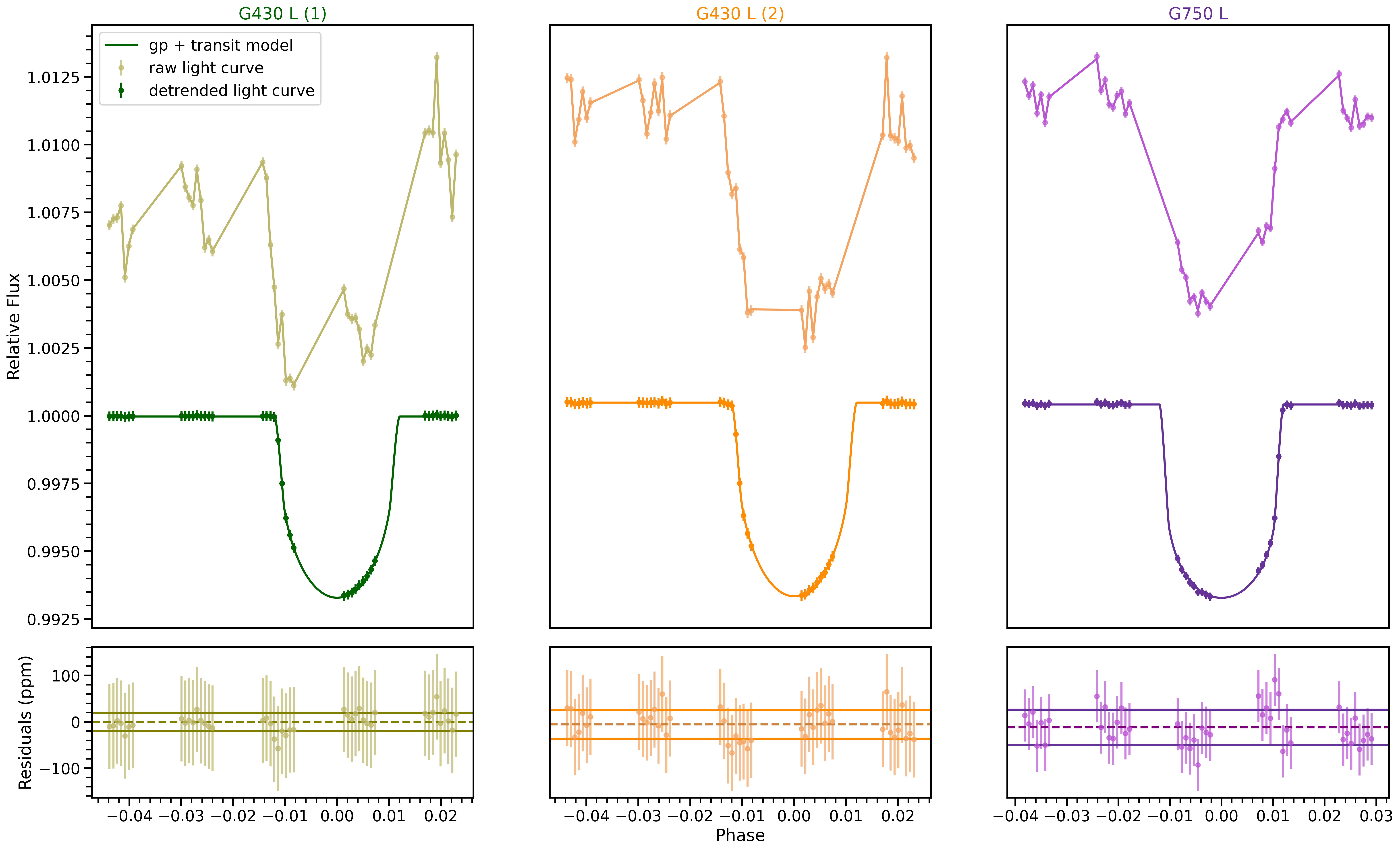}
    \caption{\textit{Top:} The raw and and detrended white light curves for each HST visit in the STIS G430L (green and orange) and STIS G750L gratings (purple). The best-fit light curve model is overplotted. \textit{Bottom:} Corresponding residuals from the white light curve fit are shown in the colored circles. The solid lines correspond to one times the standard deviation.}
    \label{fig:1}
\end{figure*}

We use a Gaussian process (GP) model over a linear model to handle the systematics as it was concluded in \cite{Allen_2024}, where the choice of detrending method does not significantly impact the transmission spectrum's shape although it affects the uncertainties. To prevent overfitting, we apply an exponential prior to the GP detrending vector coefficients which heavily penalizes unfavorable regressors. Starting with the 10 jitter vectors mentioned before, we perform principal component analysis (PCA) to generate a set of vectors that are not correlated with each other. We use them as an initial set of input parameters aiming to reduce the dimensionality of the fit with the expectation of the exponential priors. First, we re-normalize the principal components (PCs) and test the GP detrending method with the PCs ranging from 1 to 10 as input vectors. Upon completion of testing, we found that the optimal systematics model for the G430L and G750L data is eight PCs. Implementing the best systematics model, we perform a white light curve fit as seen in \autoref{fig:1} and spectroscopic fits in \autoref{fig:2} and \autoref{fig:3}.

\subsection{White Light Curve Fitting} \label{sec:white light}

After extracting the spectra and cleaning it, we start the analysis of the time-series light curves. To obtain a white light curve, we sum the flux over the entire wavelength range, from 0.29 to 0.57 $\micron$ for G430L and from 0.53 to 1.02 $\micron$ for G750L, for each exposure.  We convert the exposure time stamps given in JD$_{\mathrm{UTC}}$ to BJD$_{\mathrm{TBD}}$ using \textit{barycorrpy} \citep{Kanodia_2018}. We normalize each white light curve with respect to the median flux value of the out of transit data in a visit, and for the light curve fits we use \textit{juliet} \citep{juliet}. The modeled flux measurement consisted of a combination of a geometric transit model with limb darkening, and a systematic error correction model, the latter described in Section \ref{sec:systematics}. In this study, we adopt the stellar and planetary parameters from \cite{Hartman_2011}, as several parameters in \cite{Wakeford_2017_2} are derived from \cite{Hartman_2011}. For the remaining parameters not included in \cite{Hartman_2011}, we directly use those from \cite{Wakeford_2017_2}. This approach ensures consistency and allows for a meaningful comparison with the findings of \cite{Wakeford_2017_2} (see \autoref{table 1}). 
We used an exponential prior for the detrending vectors with the standard values from the SHEL pipeline. In the case of a resulting poor fit to the data, we tested using a range of exponential priors, a range of priors for the \textit{sigma\_w} (a jitter added in quadrature to the error bars of the instrument) and \textit{GP\_sigma} (amplitude of the GP) parameters to improve the fits which was accomplished without affecting the resulting \textit{$R_p/R_s$} and its error. During the transit fit, we fix all of the system parameters except \textit{$R_p/R_s$}, we fit for all the coefficients of the systematics detrending vectors and additionally fit for \textit{mflux} (the offset relative flux for the photometric instrument) and \textit{sigma\_w} which are inherent parameters from \textit{juliet}. When dealing with limb darkening, we fix the values to the calculation from the theory using \textit{ExoTiC-LD} \citep{Grant_2024} where we use the Kurucz stellar grid \citep{Kurucz_1993}, we assume T$_{eff}= 5079\pm$88, [M/H] = -0.04$\pm$0.08 and log g$_{\star}=4.56\pm0.06$ \citep{Hartman_2011} and use the quadratic limb darkening law parametrization from Kipping of q$_1$ and q$_2$ \citep{Kipping_2013}. The SHEL pipeline has several options for the nested sampling under which we select \textit{dynamic dynesty} for the posterior sampling of the parameter errors. 

\subsection{Spectral Light Curve Fitting} \label{sec:spectroscopic}

To produce spectroscopic light curves for the G430L observations, we bin the data into four spectrophotometric channels between 0.29 $-$ 0.57 \micron. The binning scheme was determined by balancing signal-to-noise ratio (SNR) and spectral resolution, notably the spectra in the bluer end have lower counts than in the redder part. As a result, the SNR across bins ranges from approximately 17,000 to 48,000 for both visits. We use continuous binning to avoid stellar absorption lines; by carefully selecting bin widths to avoid known stellar lines, the method reduces systematic noise and improves SNR. Proper binning strategies enhance the reliability of spectral analysis by minimizing stellar interference. This approach is essential for accurately retrieving exoplanet atmospheric properties, particularly in cases where stellar absorption lines overlap with planetary features. Using the results and testing from the white light curve fit, we carry out the same fits for the spectroscopic light curves where we fix the system parameters minus \textit{$R_p/R_s$}, fit for the coefficients of the systematics detrending vectors and additionally fit for \textit{mflux} and \textit{sigma\_w}. For limb darkening, we calculate the coefficients using \textit{ExoTiC-LD} for each bin and fix the values. 

In the case of G750L, we used a similar bin configuration as \cite{Wakeford_2017_2} as we wanted to achieve a consistent reduction to compare with the previous published results. We thus binned the G750L spectra into seven spectral bands between 0.53 $-$ 1.02 \micron. Following that, we performed the same fitting approach as for the G430L observations discussed above.

\section{Results} \label{sec:Results}

\begin{figure*}[ht] 
    \centering
    \includegraphics[width=1\textwidth]{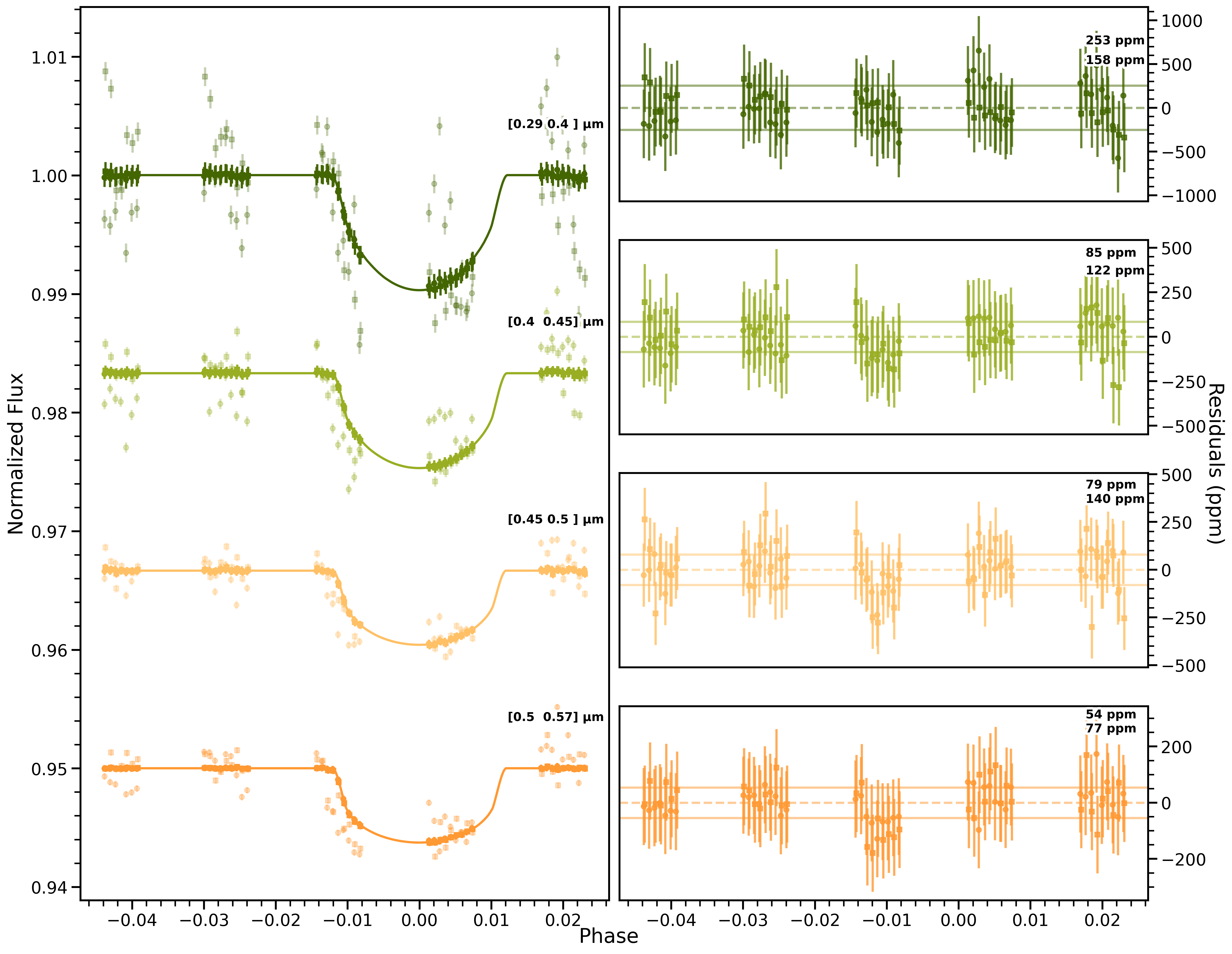} 
    \caption{Spectroscopic light curve fits for HST STIS G430L (visit 13 and visit 14, phased together). Left panel: Detrended light curves (darker points), raw data from visit 13 (circles) and visit 14 (squares), and the best-fit transit model (solid lines). Each wavelength bin is offset vertically by an arbitrary constant for clarity and the binning wavelength range is shown above the post-transit baseline. Right panel: Corresponding residuals in parts per million (ppm). The standard deviation of the residuals is reported in the residuals plot, we report two values which correspond to the two different visits.}
    \label{fig:2}
\end{figure*}

\begin{figure*}
    \centering
    \includegraphics[width=1\textwidth]{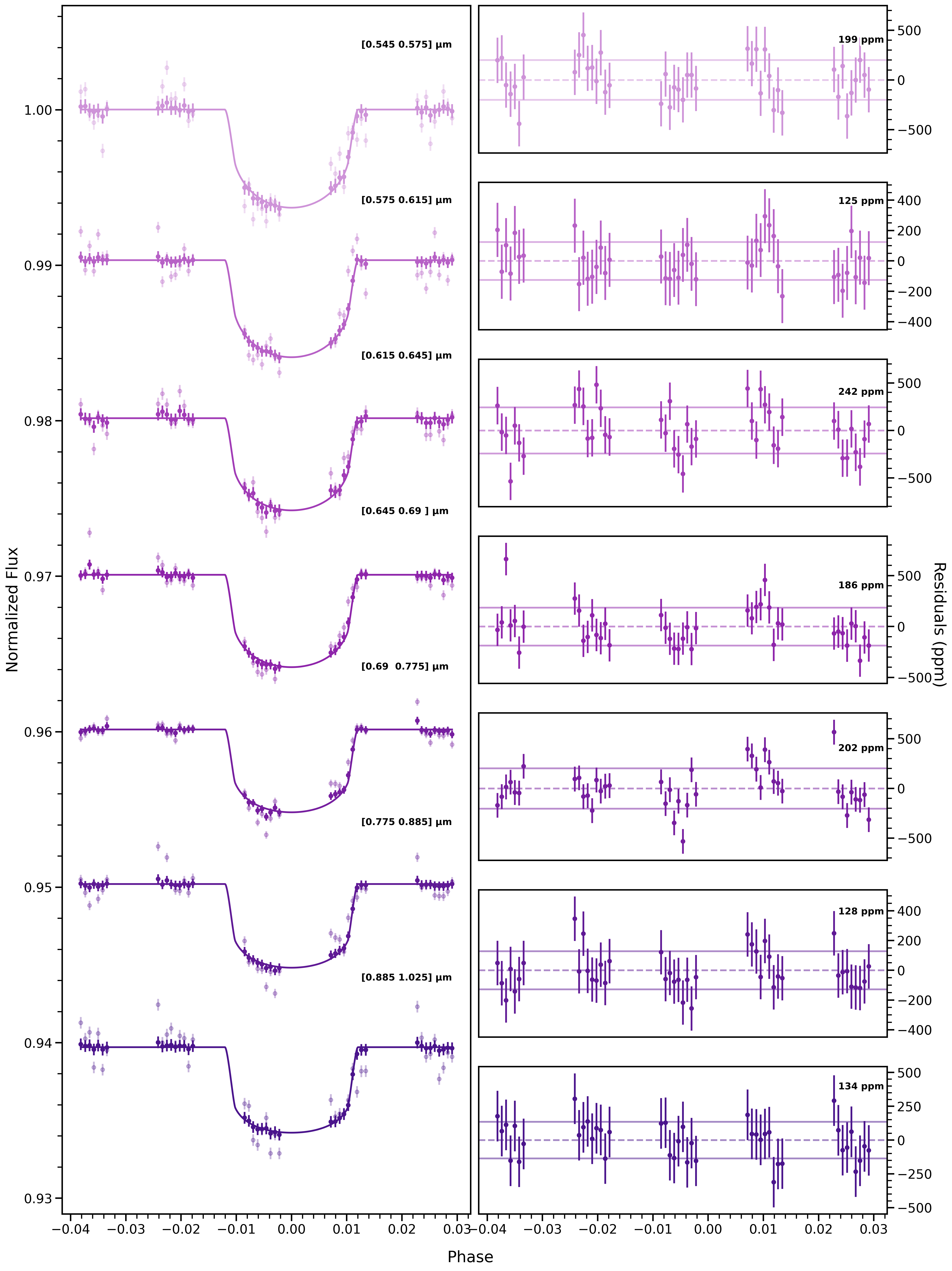} 
    \caption{Same as \autoref{fig:2}, but for one visit with the G750L grating.}
    \label{fig:3}
\end{figure*}

The results of our white light curve analysis of HAT-P-26b are listed in \autoref{table 2}. The raw, systematics-corrected light curves and the best-fit transit models are shown in \autoref{fig:1}. We note that when comparing the two transit depths from the two STIS G430L visits, the values mostly agree with each other to within 1.5$\sigma$, indicating that there is not significant stellar photometric variability between the visits and HAT-P-26 is considered to be a quiet star \citep{Wakeford_2017_2}.

\begin{table} [h]
    \caption{White Light Curve Derived Radii}
    \hspace*{-1.5cm}
    \centering
    \begin{tabular}{c c c}
    \hline\hline
    Instrument & Wavelength (\micron) & \textit{$R_p/R_s$}\\
    STIS G430L (1) & 0.29 $-$ 0.57 & 0.0679$_{-0.0042}^{+0.0048}$\\
    STIS G430L (2) & 0.29 $-$ 0.57 & 0.0758$_{-0.0022}^{+0.0022}$\\
    STIS G750L & 0.53 $-$ 1.02 & 0.0700$_{-0.0017}^{+0.0015}$\\
    \hline
    \end{tabular}
    \label{table 2}
\end{table}

\begin{figure}[h]
\hspace{-0.79cm}
    \includegraphics[width=0.53 \textwidth]{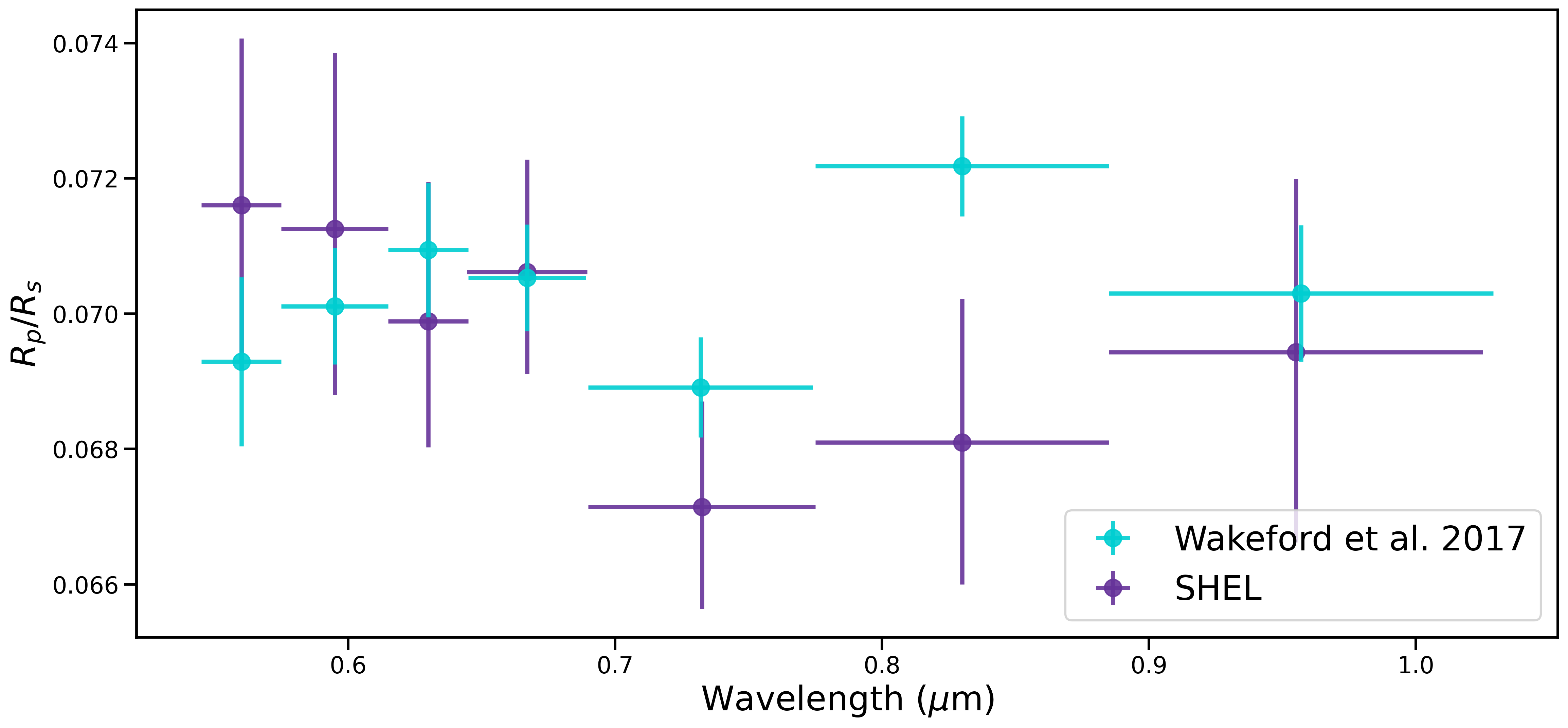}
    \caption{Transmission spectrum comparison of HAT-P-26b from HST STIS G750L grating. The data points shown in purple correspond to the work done here and the ones in light blue to the analysis from \cite{Wakeford_2017_2}.}
    \label{fig:4}
\end{figure}

We compare our re-analyzed transmission spectrum of HAT-P-26b using HST G750L with the previously published observations in \cite{Wakeford_2017_2} in \autoref{fig:4}. We can see that there is an agreement between the two reductions with the exception of one data point in the region between 0.8 $-$ 0.9 \micron. The discrepancies likely arise because the pipeline used in this analysis implements GPs while the published one uses a Lavenberg-Marquart (L-M) least-squares algorithm.

We construct an extended transmission spectrum for HAT-P-26b covering from the optical to infrared, 0.29 $-$ 5.0 \micron, by combining STIS, WFC3 and Spitzer observations where the WFC3 and Spitzer points come from \citet{Wakeford_2017_2}. \autoref{table 3} shows the measured planet-star radius ratios from the spectroscopic light curves analysis with the bin configuration for each grating. For the case of G430L, we use a joint fit function that takes into account the two corresponding visits and outputs a combined \textit{$R_p/R_s$} with its respective error. With multiple visits, \textit{juliet} assumes that the system parameters are the same across both visits and only one posterior distribution is assumed for \textit{$R_p/R_s$}.  This parameter is generated based on the combined data and the errors are given by the standard deviation of the posterior distribution.

\begin{figure*}[ht]
    \centering
    \includegraphics[width=1 \textwidth]{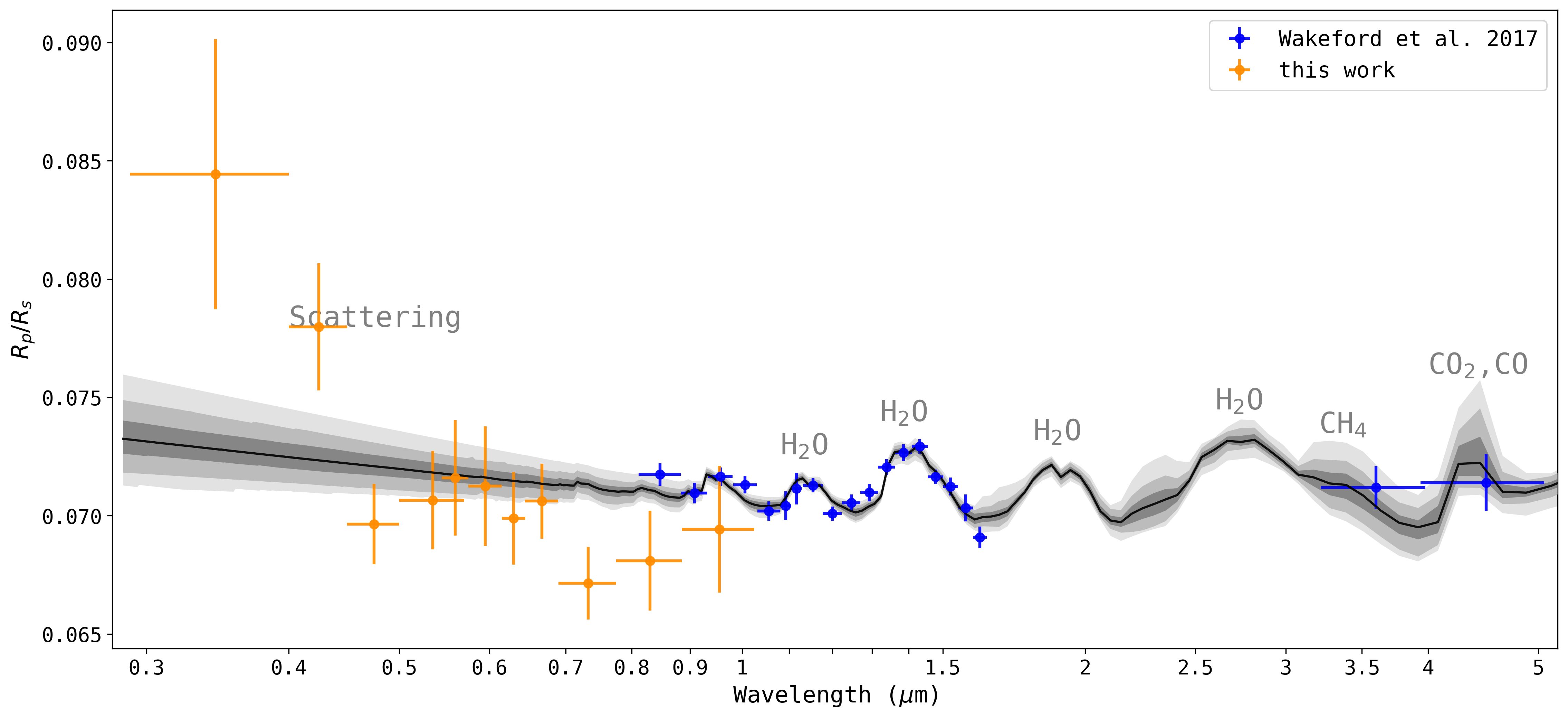}
    \caption{Broadband transmission spectrum of HAT-P-26b from HST STIS + WFC3 + Spitzer IRAC with the best-fitting model (black line) derived using the retrieval code ATMO. The work presented here is colored in orange using STIS G430L and G750L while the data from \cite{Wakeford_2017_2} is shown in blue corresponding to WFC3 G102 and G141, and Spitzer IRAC. The best-fitting model is cloud-free and includes haze opacity. We show the results for model with hazes only with 1$\sigma$, 2$\sigma$ and 3$\sigma$ uncertainty in the dark to light gray shaded regions.}
    \label{fig:5}
\end{figure*}

The transmission spectrum is shown in \autoref{fig:5} in comparison with a retrieval atmospheric model (to be detailed in \autoref{retrievals}). A compilation of the spectroscopic light curve channels with the corresponding best-fit transit models is provided in \autoref{fig:2} and \autoref{fig:3}. Looking at \autoref{fig:5}, we see that the transmission spectrum shows a moderate slope in the blue optical, suggesting Rayleigh-like scattering from aerosols.

\begin{table}
\caption{Spectroscopic Light Curve Fit Results}
    \begin{tabular}{c c}
    \hline\hline
    Wavelength (\micron) & $R_p/R_s$\\ 
    \hline
    \multicolumn{2}{c}{STIS G430L}\\
    \hline
    0.29 $-$ 0.40 & 0.0844 $\pm$ 0.0057\\ 
    0.40 $-$ 0.45 & 0.0780 $\pm$ 0.0027\\ 
    0.45 $-$ 0.50 & 0.0700 $\pm$ 0.0017\\ 
    0.50 $-$ 0.57 & 0.0707 $\pm$ 0.0021\\ 
    \hline
    \multicolumn{2}{c}{STIS G750L}\\
    \hline
    0.53 $-$ 0.58 & 0.0716 $\pm$ 0.0024\\ 
    0.58 $-$ 0.62 & 0.0713 $\pm$ 0.0025\\ 
    0.62 $-$ 0.65 & 0.0699 $\pm$ 0.0020\\ 
    0.65 $-$ 0.69 & 0.0706 $\pm$ 0.0016\\ 
    0.69 $-$ 0.78 & 0.0671 $\pm$ 0.0015\\ 
    0.78 $-$ 0.89 & 0.0681 $\pm$ 0.0021\\ 
    0.90 $-$ 1.02 & 0.0694 $\pm$ 0.0027\\ 
    \hline
    \end{tabular}
    \label{table 3}
\end{table}

\subsection{Atmospheric Retrieval: ATMO} \label{retrievals}

To investigate the atmospheric properties of HAT-P-26b, we used the 1-D radiative-convective equilibrium model ATMO \citep{Tremblin_2015,Tremblin_2016, Goyal_2018} which has been widely used as a spectroscopic retrieval model for planetary atmospheres in both transmission and emission spectroscopy \citep{Wakeford_2017_2,Evans_2017,Alam_2018,Rathcke_2021,Fu_2022}. With the new STIS data, we focused our model comparisons on differing cloud properties. We do not test other parameters as there is a clear lack of Na or K absorption features in the measured spectrum \citep{Fairman_2024}. For the chemistry, we use a free-element chemical equilibrium approach where we let C and O (C/C$_\sun$ and O/O$_\sun$) to be free with the remaining elements to be varied with a single trace metallicity parameter ($Z_{tr}/Z_\sun$). In our spectrum, water is the only prevalent molecular feature making O/O$_\sun$ the best constraining parameter. For this scenario, oxygen is a proxy for the metallicity of the planet.

The ATMO retrieval model incorporates a relatively straightforward approach to modeling clouds and hazes, and does not account for the distribution of aerosol particles \citep{Goyal_2018}. The scattering from aerosol ``haze" is represented as enhanced Rayleigh-like scattering. For the condensate ``cloud" absorption, it is assumed to exhibit a gray wavelength dependence. ATMO implements an aerosol ``haze" scattering with $\delta_{haze}$, an empirical enhancement factor, the scattering cross section of molecular hydrogen at 0.35 $\micron$ ($\sigma_{0}$) and a factor that determines the wavelength dependence ($\alpha_{haze}$). For the ``cloud" absorption, it uses an empirical value for the strength of the gray scattering ($\delta_{cloud}$) and the scattering opacity due to molecular hydrogen at 0.35 $\micron$ ($\kappa_{H2}$). These parameters and their bounds help describe the effects of clouds and hazes in the atmospheric model implemented in the retrieval. To fit the data we consider four different atmospheric retrieval models: clouds + hazes, clouds only, hazes only and no clouds or hazes. For each case, we assumed an isothermal temperature structure based on the conclusions in \cite{Wakeford_2017_2} and we fit for temperature, planetary radius,  cloud and haze opacity (depending on the case) and for the abundances of H$_2$O \citep{Barber_2006}, CO$_2$ \citep{Tashkun_2011}, CO \citep{Rothman_2010}, CH$_4$ \citep{Yurchenko_2014}, NH$_3$ \citep{Yurchenko_2011}, H$_2$S \citep{Rothman_2013}, HCN, C$_2$H$_2$ and SO$_2$ \citep{Underwood_2016}(see \cite{Goyal_2018,Goyal_2020} for full description).

\begin{figure*}
    \centering
    \includegraphics[width=1 \textwidth]{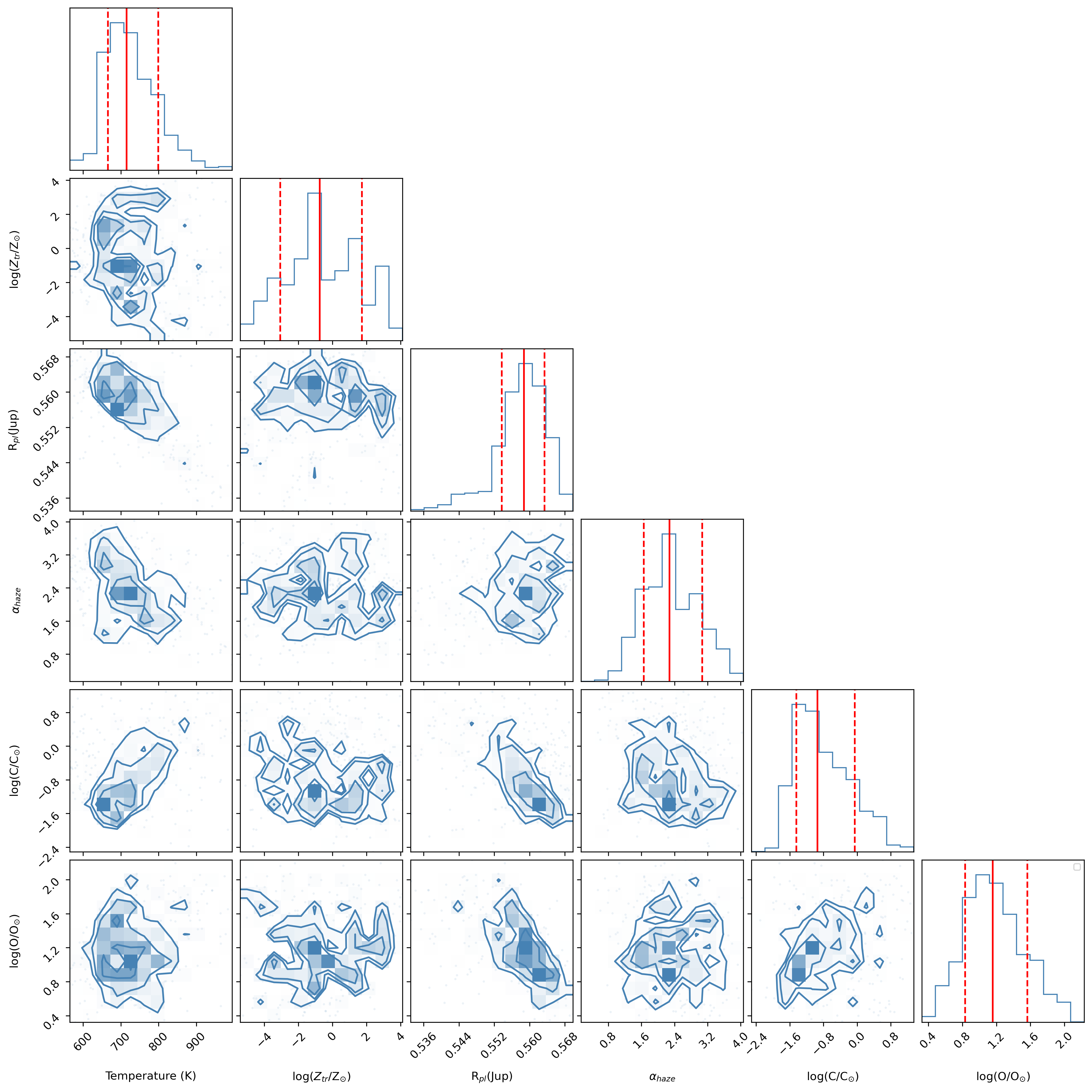}
    \caption{HAT-P-26b retrieval posterior showing the distribution of the retrieved parameters from the HST + Spitzer transmission spectrum using ATMO for the best-fitting model (haze-only scenario). Contours corresponding to 1, 1.5 and 2$\sigma$ are shown (blue). The 1D histograms show the parameters' mean value and 1$\sigma$ range (red).}
    \label{fig:6}
\end{figure*}

The results of our ATMO retrieval fits are shown in \autoref{table 4}. To determine relevant parameters such as metallicity, temperature, C/O ratio and $\alpha_{haze}$, we employ Bayesian Model Averaging (BMA) by combining the posterior distributions of all models, with each model's contribution weighted according to its Bayesian evidence following \cite{Wakeford_2016,Wakeford_2017_2}. The Bayesian evidence for a given model (${\mathit{E_q}}$) is linked to the Bayesian Information Criterion (BIC) through an approximation of the marginal likelihood, expressed as \begin{equation}
\ln E_q = \ln P(D|S_q) \approx -\frac{1}{2} BIC\end{equation} To compare different retrieval models, the evidence values are converted into weights (${\mathit{W_q}}$) which represent the relative probability of each model. These weights are determined by normalizing the evidence values across all models as \begin{equation}
W_q = \frac{E_q}{\sum\limits_{q=0}^{N_q} E_q}
\end{equation} The weighted mean ($\alpha_{m}$) of a parameter ($\alpha_{q}$) is computed using these weights to account for contributions from different models \begin{equation}
\alpha_m = \sum\limits_{q=0}^{N_q} (W_q \times \alpha_q)\end{equation} The overall uncertainty, $\sigma(\alpha)$, \begin{equation}
\sigma(\alpha) = \sqrt{\sum\limits_{q=0}^{N_q} W_q \left[ (\alpha_q - \alpha_m)^2 + \sigma_{\alpha_q}^2 \right]}
\end{equation} incorporates both the variance of individual parameter values around the mean and their associated uncertainties. This approach ensures that the final uncertainty reflects both the spread of parameter estimates and the confidence in each model's contribution. Our retrieval analysis finds a 15$_{-8}^{+22}$$\times$ solar metallicity. The retrieved slope parameter ($\alpha_{haze}$) exhibits a small increase when clouds are included in the model. However, given the propagated uncertainties, this difference is not statistically significant. The confidence interval overlap indicates that the slopes from both models are consistent within error. Consequently, we find no strong evidence that the inclusion of clouds significantly alters the retrieved haze slope for this exoplanet. The corner plot in \autoref{fig:6} shows the distributions of each parameter of interest.

\begin{table*}[ht]
    \centering
    \renewcommand{\arraystretch}{1.2} 
    \setlength{\tabcolsep}{6pt} 
    \caption{Atmospheric retrieval configuration scenarios with their respective results.}
    \label{table 4}
    \hspace{-2cm}
    \begin{tabular}{lccccccc}
        \toprule
        \midrule
        Model & Hazes Only & Clouds + Hazes & Clouds Only & No Clouds/Hazes & \textbf{BMA}\\
        \midrule
        $\chi^2$ & 40.16 & 40.58 & 51.80 & 75.46 \\
        $\chi^2_\nu$ & 1.67 & 1.76 & 2.07 & 2.90 \\
        BIC & 62.41 & 68.05 & 71.11 & 91.74 \\
        $\mathit{W_q}$ & 0.93 & 0.056 & 0.012 & $2.0\times10^{-8}$ \\
        \textit{Temperature} (K) & $714^{+85}_{-49}$ & $730^{+76}_{-73}$ & $703^{+72}_{-62}$ & $896^{+84}_{-51}$ & \boldmath$715^{+84}_{-51}$\\
        $\log[(Z_{tr}/Z_\odot)]$ & $-0.75_{-2.27}^{+2.49}$ & $-1.17_{-2.19}^{+2.43}$ & $-0.05_{-2.58}^{+2.57}$ & $-0.53_{-3.45}^{+3.15}$ & \boldmath$-0.76_{-2.27}^{+2.49}$ \\
        $\textit{$R_{pl}$}$(Jup) & $0.5587_{-0.0050}^{+0.0047}$ & $0.5589_{-0.0049}^{+0.0050}$ & $0.5555_{-0.0052}^{+0.0057}$ & $0.5530_{-0.0014}^{+0.0012}$ & \boldmath$0.5600_{-0.0050}^{+0.0047}$ \\
        $\alpha_{haze}$ & $2.3_{-0.6}^{+0.8}$ & 2.5$_{-0.7}^{+0.7}$ & $-$ & $-$ & \boldmath$2.3_{-0.6}^{+0.8}$\\
        $\log[(O/O_\odot)]$ & $1.2^{+0.4}_{-0.3}$ & $1.1^{+0.4}_{-0.5}$ & $1.2^{+0.4}_{-0.3}$ & $2.5^{+0.1}_{-0.1}$ & \boldmath$1.2^{+0.4}_{-0.3}$\\
        \textit{O/O$_\odot$} ($\times$ solar) & $14^{+22}_{-8}$ & $13^{+17}_{-9}$ & $17^{+23}_{-9}$ & $288^{+115}_{-70}$ & \boldmath$15^{+22}_{-8}$\\
        \textit{C/O} & $0.004^{+0.020}_{-0.003}$ & $0.005^{+0.003}_{-0.003}$ & $0.003^{+0.006}_{-0.002}$ & $0.0004^{+0.0060}_{-0.0004}$ & \boldmath$0.004_{-0.003}^{+0.020}$\\
        $\textit{$P_{cloud}$}$(mbar) & $-$ & $0.48^{+0.07}_{-0.07}$ & $0.24^{+0.07}_{-0.07}$ & $-$ & \boldmath$0.44^{+0.12}_{-0.12}$\\
        \midrule 
        \bottomrule
    \end{tabular}
    \vspace{0.5em} 
    \begin{center}
    \textit{Note}: The number of data points (N) used were 31 for all cases. For the number of parameters (k), we used 8, 6, 7, and 5, listed in the same order as the models in the table. The degrees of freedom (DOF) for each model were 24, 23, 25, and 26, following the same order.
    \end{center}
    
\end{table*}

\section{Discussion} \label{sec:Discussion}

We interpret the optical to infrared transmission spectrum of HAT-P-26b in the context of the observed mass-metallicity relationship in exoplanetary atmospheres. Our retrieval of the 0.29 $-$ 5.0 $\micron$ HST + Spitzer spectrum indicates the presence of a scattering slope and water vapor (H$_2$O), the latter initially reported in \cite{Wakeford_2017_2}. With our new wavelength coverage in the blue optical we explored previous predictions of an absorbing cloud deck (\cite{Wakeford_2017_2}) which we do not find but rather a scattering slope best explained by haze-dominated scenario. Considering the haze properties, we find that the scattering alpha ($\alpha_{haze}= 2.3_{-0.6}^{+0.8}$) indicates a weaker wavelength-dependence than Rayleigh scattering ($\alpha_{haze}= 4$) which suggests absorbing rather than purely scattering particles, this can be caused by larger particles,  specific particle compositions or a mix of particle sizes such as from settling grains \citep{Pont_2013}. 

An essential objective of exoplanet atmospheric studies is to connect atmospheric properties to planet formation history, to better understand the evolution of close-in giant planets from their formation to current day locations and the impact of growth and migration on their atmospheric properties. Metallicity measurements provide a crucial link to planet formation mechanisms. With the metallicity derived using ATMO (15$_{-8}^{+22}$$\times$ solar) we observe that HAT-P-26b lies below the expected mass-metallicity trend for exoplanets when compared to our Solar System's gas giants ($\sim80\pm20 \times$ solar for Neptune), placing it $\sim 3 \sigma$ below the trend. If we compare our result with what was previously derived in \cite{Wakeford_2017_2} (4.8$_{-4.0}^{+21.5}$$\times$ solar) and \cite{MacDonald_2019} (18.1$_{-11.3}^{+25.9}$$\times$ solar), our result is in agreement to $< 1\sigma$ for both cases. The overall conclusion from our result is consistent with the previous findings described in \cite{Wakeford_2017_2}, where its formation scenario is aligned with envelope accretion models \citep{Lee_2016}. This scenario argues that the majority of hot Neptunes acquire their envelopes in situ, just before their protoplanetary disks disappear. 

We note that the addition of the optical wavelength covered by the STIS G430L grating helps provide constraints on the metallicity value and most importantly, insights and constraints on the aerosol properties \citep{Sing_2011,Pont_2013,Rustamkulov_2023,Fu_2024, Fairman_2024}.

\section{Conclusions} \label{sec:Conclusion}
We presented new HST G430L (0.29 $-$ 0.57$\micron$) transmission spectra and a re-reduction of the HST STIS G750L (0.53 $-$ 1.02 $\micron$) previously published data along with previous observations from HST WFC3 G102 and G141, and Spitzer of the warm Neptune HAT-P-26b. The main findings are summarized below:

\begin{enumerate}
    \item The transmission spectrum of HAT-P-26b is characterized by an optical scattering slope that is less wavelength-dependent than Rayleigh scattering, indicating larger particles or a mix of particle sizes. 
    \item We retrieve the planet's atmospheric properties using ATMO. The result for metallicity, 15$_{-8}^{+22}$$\times$solar, is in agreement with the previously published result of HAT-P-26b from \cite{Wakeford_2017_2} and \cite{MacDonald_2019}.
\end{enumerate}

 HAT-P-26b has been observed as part of the JWST Cycle 1 Guaranteed Time Observation (GTO) programs (GTO-1312, PI: N. Lewis) with multiple instruments (e.g., NIRISS SOSS and NIRSpec G395H) which will provide further constraints on the atmospheric composition and structure. Complementing the new JWST observations with the existing HAT-P-26b data including the HST STIS G430L analysis presented here, significantly enhances the ability to characterize the planet’s atmosphere. The inclusion of STIS G430L extends the spectral coverage into the visible range, enabling improved constraints on optical scattering slopes and haze properties.  This data serves as a critical baseline for interpreting molecular absorption characteristics observed in the infrared by WFC3, Spitzer, and JWST, ensuring consistency and accuracy across the entire spectrum. By integrating these datasets, a more comprehensive and precise understanding of HAT-P-26b’s atmospheric properties and formation history can be achieved.
 
\software{astropy \citep{astropy_2013, astropy_2018}, ATMO \citep{Tremblin_2015,Tremblin_2016,Goyal_2018}, barycorrpy \citep{Kanodia_2018}, batman \citep{batman}, corner \citep{corner}, dynesty \citep{dynesty}, george \citep{george}, juliet \citep{juliet}, jupyter \citep{jupyter}, matplotlib \citep{matplotlib}, multinest \citep{Feroz_2008,Feroz_2009, Feroz_2019}, NumPy \citep{numpy}, pandas \citep{pandas,pandas_2}, SciPy \citep{scipy}, transitspectroscopy \citep{transitspectroscopy}}. 

\section*{Acknowledgments}
This work is based on observations made with the NASA/ ESA Hubble Space Telescope obtained at the Space Telescope Science Institute (STSci), which is operated by the Association of Universities for Research in Astronomy, Inc. The observations associated with this analysis are the HST-GO programs 14767 and 14110 and the datasets can be accessed via this \dataset[DOI]{https://doi.org/10.17909/ztrs-da52}. Support for this work was provided by NASA through grants under the HST GO-14767 and HST AR-6634 program from STScI. The authors thank N. Espinoza for helpful discussions. 

\bibliography{main_arxiv}

\end{document}